\documentclass[twocolumn]{aastex701}

\usepackage[utf8]{inputenc}

\usepackage{makecell}
\usepackage{tabularx}

\usepackage[starfontserif]{starfont}
\usepackage{amsmath}
\DeclareSymbolFont{starfontsym}{OT1}{sts}{m}{n}
\DeclareMathSymbol{\mathJupiter}{\mathord}{starfontsym}{106}

\usepackage{graphicx}
\usepackage{hyperref}

\begin{document}

\title{On the origin and dynamical evolution of Jupiter's moon Amalthea}
\author[0000-0002-2478-3084]{Ian R. Brunton}
\affiliation{Division of Geological and Planetary Sciences, California Institute of Technology}
\email[show]{\href{mailto:brunton@caltech.edu}{brunton@caltech.edu}}

\author[0000-0002-7094-7908]{Konstantin Batygin}
\affiliation{Division of Geological and Planetary Sciences, California Institute of Technology}
\email{kbatygin@caltech.edu}

\begin{abstract}

Interior to the orbits of Jupiter's iconic Galilean moons are four small satellites with individual mean radii, $R\lesssim 84$ km. Multiple lines of evidence suggest that these bodies formed at a more distant location in Jupiter's circumplanetary disk before coming to reside at their current short-period orbits. Nonetheless, how these moons dynamically evolved to such a location has yet to be explained in the emerging paradigm of Jovian satellite formation. Here, we present a quantitative model for the origin of the largest of these inner moons, Amalthea, that can be extended to its neighbor, Thebe, and to other small bodies in astrophysical disks. We propose that Amalthea's anomalous features are due to it having formed alongside the Galileans in a reservoir of satellitesimals located at a large jovian-centric distance. As the innermost Galilean, Io, migrated inward from this reservoir, it captured the satellitesimal, Amalthea, into resonance and shepherded the small body to its modern neighborhood. During migration through the disk, dissipative forcing from aerodynamic drag induces overstable librations in the Io-Amalthea resonance, such that only a narrow range of nebular parameters can accommodate the requisite long-range transport. In particular, the disk-aspect ratio, $h/r$, emerges as the key variable. Our calculations indicate that the circumjovian disk had a scale height of at least $h/r\gtrsim0.08$, implying a relatively hot, actively accreting disk during the epoch of satellite formation. These results thus shed light on the evolution of the Jovian system, along with the more general phenomenon of satellite-disk interactions.  

\end{abstract}

\section{Introduction}

Amalthea, the fifth largest moon in the Jovian system, is an irregularly-shaped object of mean radius, $R=$ 84 km, revolving around Jupiter on a short-period orbit of just $\sim$12 hours. Nestled interior of the far larger Galileans, its proximity to Jupiter combined with the moon's dark, reddened surface, proved it near impossible for early astronomers to spot amidst the brilliant glare of its giant host and iconic neighbors. But almost three centuries after Galileo's discovery of the major Jovian satellites, the sharp-eyed astronomer, E. E. Barnard, sighted Amalthea with the 36-inch telescope at the Lick Observatory in California \citep{Barnard92}. Chiming in on the debates of Amalthea's origins, Barnard remarked, ``there is no question this satellite has been there all along," seeing that his careful measurements ``show that its orbit lies sensibly in the plane of Jupiter's equator and that consequently the satellite is not a new addition to the Jovian family" \citep{Barnard93}. 

Prior to a proper measurement of Amalthea's mass and volume, Barnard's conviction was not far from the prevailing thought, in which an \textit{in situ} formation for Amalthea would imply a relatively dense, refractory composition -- in line with the trend of its Galilean neighbors \citep{Gradie+80,Lunine&Stevenson82,Greenberg10}. However, while analyzing flyby data from NASA's \textit{Galileo} space probe, \citet{Anderson+02,Anderson+05} determined the bulk density of Amalthea to be 0.857 $\pm$ 0.099 g/cm$^3$; a remarkably low value, suggesting it a porous body composed mainly of water ice and rock. Further complementing these findings, \citet{Takato+04} obtained near-IR spectra of Amalthea's trailing side, which reflected absorption features characteristic of hydrous materials at the surface, and a close resemblance to the most distant Galilean moon, Callisto. These data collectively point to Amalthea having formed in a more distant, cooler region of the circumjovian nebula, where it then must have been dynamically transported to its modern location \citep{Takato+04, Anderson+05}.\footnote{Both \citet{Takato+04}, and \citet{Anderson+05} -- along with others (e.g., \citealt{Prentice&terHaar79,Prentice03}) -- also speculated that Amalthea may be a captured object originating from a heliocentric orbit. We address the faults of this hypothesis in the appendix.}

In the time since these observations, substantial theoretical advancements have been made with respect to Jovian satellite formation, in which the emerging paradigm has primarily focused on understanding the conglomeration and orbital architecture of the Galilean moons \citep[e.g.,][]{CanupWard06,Shibaike+19,RonnetJohansen20,BatyginMorbidelli20}. Not yet properly formulated in this picture, however, is a quantitative model for the origin of the four non-Galilean regular satellites of the Jovian system: the Amalthea group, so named for their largest member (see Table \ref{tab:moonprops}). 

\begin{table}
\label{tab:moonprops}
\begin{center}
The inner moons of Jupiter today
\setlength{\tabcolsep}{3.8pt}
\begin{tabular}{c|cccccc}
\hline\hline
     & $m$ [kg] & $R$ [km] & $\bar\rho$ [$\frac{\text{kg}}{\text{m}^3}$] & $a$ [$R_\text{J}$] &  $e$ & $i$ [$^\circ$]\\
     \hline
    \textbf{Io} & 9$\times$10$^{22}$ & 1800 & 3500 & 5.91  & .004 & .05\\
    Thebe & --- & 49 & --- & 3.11 & .02 & 1.08\\
    \textbf{Amalthea} & 2$\times$10$^{18}$ & 84 & 860 & 2.54 & .003 & .37\\
    Adrastea & --- & 8.2 & --- & 1.81  & .002 & .03\\
    Metis & --- & 22 & --- & 1.79 & .0002 & .06\\
    \hline
\end{tabular}
\caption{Io and the Amalthea group, listing mass, $m$, mean radius, $R$, and mean density, $\bar\rho$, with relevant orbital parameters of semi-major axis, $a$, eccentricity, $e$, and inclination, $i$. Modern Jupiter radius, $R_\text{J} \approx$ 71500 km. The members of the Amalthea group are irregularly shaped and heavily cratered \citep{Thomas+98}. Of these four, only Amalthea's mass has been independently determined \citep{Anderson+05}.}
\end{center}
\end{table}

\subsection{This work}
\label{sub:thiswork}

In this paper, we thus put forth a model for the dynamical origins of Amalthea that fits into what we currently know of the small body, its fellow Jovian moons, and the properties of the protojovian environment. Our basic idea begins from the foundations that are generally agreed upon in all modern theories of Jovian satellite formation: the Galilean moons assembled from a reservoir of satellite building blocks located at a relatively large distance from Jupiter. As the Galileans grew in mass, their dynamic coupling with the surrounding medium eventually drove a phase of rapid inward type-I migration, upon which they subsequently anchored at the disk's inner edge (see \citealt{McKinnon23} for a review). 

\begin{figure*}
    \includegraphics[width=\textwidth]{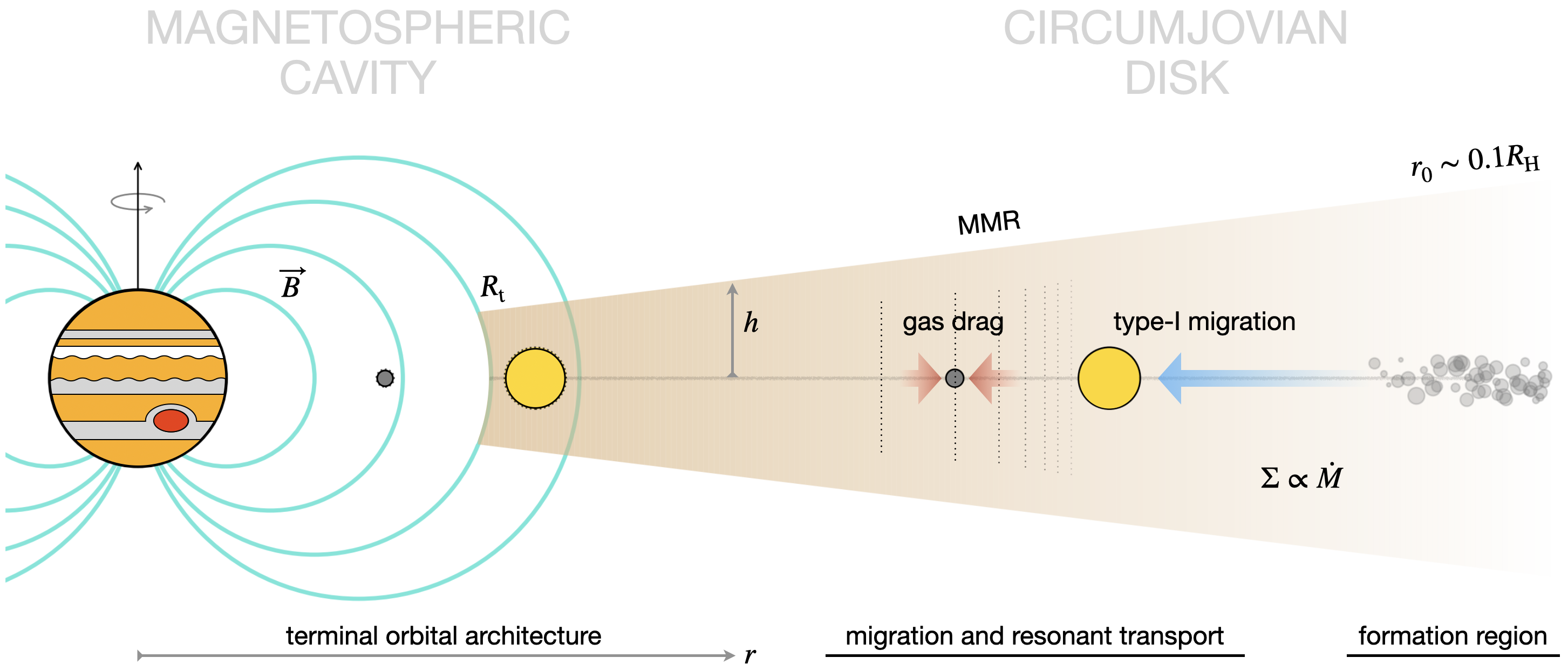}
    \caption{A general schematic of our proposed scenario for how Amalthea reached the inner Jovian system. The events take place in the last $\sim$$10^5$ years of Jupiter's disk-bearing epoch. Both Amalthea (dark gray circle) and Io (gold circle) form in a reservoir of satellite building blocks in the distant regions of the disk at approximately $r_0\sim0.1\ R_\text{H}$. Io grows large enough to commence inward type-I migration through Jupiter's circumplanetary disk of density, $\Sigma$, and scale height, $h$. The rate of Io’s orbital decay via type-I migration is far quicker than the smaller moon's decay via aerodynamic drag. Accordingly, Io captures Amalthea on an interior mean motion resonant (MMR) orbit, transporting the small moon inward to the inner Jovian system. The resonant locking is subject to overstable librations that are driven by satellite-disk interactions; principally, the disk-aspect ratio, $h/r$. Amalthea and Io must reach the magnetospheric cavity beginning at the truncation radius, $R_\text{t}$, prior to the photoevaporation of the circumjovian disk.}
    \label{fig:AmaSchem}
\end{figure*}

We consider here that a handful of small, Amalthea-like satellitesimals would have also resided on nearby orbits amidst the same reservoir of material, in which their diminutive size would limit their dynamic coupling with the gaseous disk to that of aerodynamic drag resultant from the sub-keplerian flow of the surrounding medium. At the distance of this reservoir, the characteristic timescale for orbital decay via the Galileans' type-I migration is far more rapid than any orbital decay induced via aerodynamic drag on an Amalthea-sized object, thus \textit{necessitating} that the first of the migrating Galileans, Io, would converge with the inner satellitesimal on its path inward. Io’s gravity, therefore, would have swept up this small body into an interior orbital resonance. Once locked into resonance, Amalthea would effectively hitch a ride under Io’s gravitational influence and be delivered to its short-period orbit before photoevaporation inevitably led to the demise of Jupiter's circumplanetary disk. A schematic of our proposed scenario is shown in Figure \ref{fig:AmaSchem}.

Despite the natural agreement of this scenario with Amalthea's features and the larger picture of the Galileans, the resonant transport mechanism is not so simple. Due to the large difference in mass between $m$, Amalthea, and $m'$, Io ($m/m' < 10^{-4}$), the transport of the small body succumbs to \textit{overstable librations}, whereby oscillations in the resonance build until the captured satellite breaks free \citep{Goldreich&Schlichting14,Deck&Batygin15}. The interplay between this overstability, orbital decay, and dynamical chaos arising from resonance overlap, determines whether Io can succesfully transport Amalthea and actually deliver the small moon to the inner Jovian system. Furthermore, the degree of overstability is intrinsically governed by the structure and thermodynamic properties of the circumjovian disk -- most notably, the extent of concentration of gas in the midplane characterized through the geometric aspect ratio, $h/r$. As such, while the dynamical evolution is complicated by the inherent overstability, we can exploit this complexity to investigate the properties of the disk which prove amenable to this transport mechanism, thereby revealing unique constraints on the early jovian environment. 

We begin in \S \ref{sect:Just} briefly addressing why the inner regions of Jupiter's circumplanetary disk remained inimical to the formation and/or long-term survivability of the icy, low-density Amalthea. This discussion serves to justify our methodological approach discussed in \S \ref{sect:NumExp}, and clarifies the necessity of the resonant transport mechanism. Through our simulations, we quantify the satellite-disk interactions of Amalthea and Io, and explore these relationships to uncover any constraints that emerge. Our simulations show intricate dynamical processes that deserve a proper Hamiltonian treatment which we provide in \S \ref{sect:formalism} to supplement our numerical experiments. Along the way we highlight key influences on the dynamical evolution, synthesizing this discussion in \S \ref{sect:Discussion}. Although we do focus primarily here on Amalthea, the same story may be similarly applied to the three other moons of the Amalthea group -- albeit, with important caveats that will be discussed in the final section alongside comments regarding the implications for the Jovian system and beyond.

\section{Amalthea's origins}
\label{sect:Just}

The data cited above points to Amalthea having formed in a far cooler environment than that of what could be expected so close to its accreting host. Let us examine this environment to determine a reasonable estimate for where Amalthea's journey began. 

\subsection{Satellite formation}
\label{sub:satform}

A variety of models exist for the evolution of Jupiter during the solar nebula phase of existence; models differing primarily in their accretion mechanism of the Galileans. For now, suffice to say that there is general agreement that the waning stages of Jupiter's primordial epoch -- the final $\sim$10$^5$ years of Jupiter's accretion -- correspond to the period of satellite formation and/or survivability \citep{McKinnon23}. At this time, Jupiter grew large enough to clear out its own gap within the Sun's nebula, where it maintained its own hot, relatively dense disk of material supplied by the inflow of matter from the larger solar nebula, extending out to a substantial fraction of the planet's Hill radius, $R_\text{H}$, and truncated at the inner edge, $R_t$, where effects from Jupiter's formidable magnetic field dominate the dynamics of the system. 

During this primordial stage (represented here by superscript, $\dagger$), the highly-luminous Jupiter's mean radius was larger than that of today, $R_\text{J}^\dagger \gtrsim$ 2 $R_\text{J}$; its magnetic field stronger, $B_\text{J}^\dagger \gtrsim$ 50 $B_\text{J}$; and its effective temperature higher, $T_\text{J}^\dagger \gtrsim$ 10 $T_\text{J}$ (see \citealt{Batygin&Adams25} for details). An effective temperature approaching $1500\ \rm{K}$ indicates an ambient nebular environment upwards of $750\ \rm{K}$ near the vicinity of the inner moons' modern-day orbits in even the most weakly-accreting of disks -- temperatures prohibitive to the formation of icy, low density, $R\sim100$ km objects. Furthermore, the upshot of the robust magnetic field protruding from the inflated Jupiter is that the truncation radius of the disk -- calculated as $R_t\propto (B^4R^{12})_\text{J}^{\dagger{1/7}}$ -- remains above $R_t \gtrsim 3.5$ $R_\text{J}$ throughout the relevant period of satellite formation \citep[see][]{Batygin18}. Recall that the current orbit of Amalthea (and the rest of the Amalthea group) would thus be well \textit{inside} this truncation limit (Table \ref{tab:moonprops}), and formation of any satellitesimals with radius $R \gtrsim 50$ km, \textit{inside} the magnetospheric cavity ($r < R_t$), would simply not be possible.\footnote{Even a hypothetically dense, rocky Thebe, Metis, or Adrastea (none of which yet have precise measurements for their bulk properties) could not accumulate enough mass to form inside this cavity.} 

In accounting for the above disk properties and the overall hostility of the inner disk, numerous studies have independently put forth that the Galilean satellites assembled from a reservoir of icy and rocky material in the distant regions of the nebula \citep[e.g.][etc.]{CanupWard06,Shibaike+19,RonnetJohansen20,BatyginMorbidelli20,Cilibrasi21}. The precise location of this reservoir differs primarily on the mechanism invoked for satellite formation; however, the preference for satellitesimals forming in the cooler regions of its natal environment is generic. \citet{BatyginMorbidelli20} determine this formation zone to be between $r\sim0.1$--$0.3$ $R_\text{H}$, a fiducial value that will guide the starting point for our numerical experiments in \S \ref{sect:NumExp}. But notably, models invoking alternative formation mechanisms for the satellite embryos still place this zone upwards of $r\sim0.05$ $R_\text{H}$ \citep[e.g.,][]{Shibaike+19,RonnetJohansen20}, so too necessitating some process to transport Amalthea inward.

\subsection{The circumjovian disk}
\label{sub:circdisk}

Having entered its runaway accretion phase of growth, Jupiter accretes material from the parent solar nebula at some rate, $\dot{M}$. As infall proceeds, material spins up to maintain the circumplanetary disk with a surface density, $\Sigma$, and a geometrically thin aspect ratio of $h/r\ll1$ throughout (see e.g., \citealt{Adams&Batygin25}, and the references therein). This density gradient and vertical structure govern the satellite-disk interactions, and thus the characteristic perturbations to the satellites' movement through the circumjovian nebula. To adequately account for these specific properties in N-body simulations, a common tactic is to adopt simple power-law relationships for the surface density and pressure scale height: \begin{equation}
\label{eq:Sigma}
  \begin{aligned}
    \Sigma &= \Sigma_0\left(\frac{r}{r_0}\right)^{-s}
  \end{aligned} 
\end{equation}
and 
\begin{equation}
\label{eq:h}
  \begin{aligned}
    h &= h_0\left(\frac{r}{r_0}\right)^{\beta},
  \end{aligned} 
\end{equation}
with their indices, $s$ \& $\beta$, and reference values of $\Sigma_0$ \& $h_0$ at some defined distance $r=r_0$. We too will adopt these simplifications; however, we briefly remark here on the physical meaning behind these power-law profiles to facilitate the coming discussion of overstability in the Io-Amalthea resonance. 

For a ``passive" circumplanetary disk, the dominant source of energy comes from the reprocessed radiation received directly from the hot, growing planet; whereas in an ``active" disk, the energy is derived from the heat generated within the nebula, characterized by its turbulent viscosity, $\nu$, which is parameterized by the Shakura-Sunyaev, $\alpha$ \citep{Shakura&Sunyaev73}. The balance between passive and active heating translates to a measure of $\dot{M}$ versus the planetary luminosity, and given existing contraints, even a modest $\dot{M}$ (e.g., $\sim$0.1 $
M_\text{J}/$Myr in \citealt{BatyginMorbidelli20}) is sufficient for active heating to dominate. The associated surface density of eq. (\ref{eq:Sigma}) then is ultimately linked to the accretion processes, evolving as the ratio of the exchange of material between planet and parent nebula, and the turbulence in the disk, $\Sigma \propto \dot{M}/(\alpha c_\text{s} h)$, with $c_s$ the isothermal speed of sound. 

The vertical structure of the disk is set by hydrostatic equilibrium. Assuming a locally isothermal equation of state, the gas density, $\rho$, takes on a Gaussian profile, characterized by the aspect ratio, 
\begin{equation}
\label{eq:h/r}
    \frac{h}{r} = \frac{c_\text{s}}{v_k} \propto T^{-1/2}.
\end{equation}
Notably, the mid-plane density is inversely dependent on the temperature, and is related to both $\Sigma$ and $h$,
\begin{equation}
\label{eq:rhom}
    \rho_\text{m} = \frac{\Sigma}{\sqrt{2\pi}\ h}\propto T^{-3/2}.
\end{equation} 

We emphasize these relations not only because they directly determine the strength of the satellite-disk interactions, but also because our numerical setup below is guided by established models of Jupiter's circumplanetary disk in the literature -- models that often reproduce similar outcomes for the Galilean satellites, but remain distinct through the values assigned to the processes delineated above. For example, an active, steady-state decretion disk -- most closely resembling the model in \citet{BatyginMorbidelli20} -- would be necessarily hot, with a surface density of $\Sigma_0 \approx 4000$ g/cm$^2$ at $r_0=0.1\ R_H$, falling off with $s=1.25$. In contrast, a passive, slowly-accreting disk -- resemblant of \citet{RonnetJohansen20} -- would insist on a cooler, less dense, but slightly steeper profile with $\Sigma_0 \approx 600$ g/cm$^2$ at $0.1\ R_H$ and $s=1.50$. And while both of the cited models make the common approximation of little to no flaring, with $\beta\approx0$, a warmer, active profile results in a robust thermal support structure to puff up the disk to an aspect ratio of $h/r = 0.10$; whereas the latter profile entails a more densely concentrated midplane with $h/r = 0.06$. Keeping these disk properties in mind, let's now turn to what they mean for the orbits of the satellites and their early evolution. 

\subsection{Satellite-disk interactions} 

The pressure support in the disk maintains not only the nebula's vertical structure, but also contributes to its radial structure, through $\partial P/\partial r$. This radial pressure gradient acts to ever-so-subtly alter the azimuthal velocity of the gas, so that it somewhat lags $v_k$ of the orbiting satellites \citep{Adachi+76, Weidenschilling77} -- a lag quantified by the dimensionless factor
\begin{equation}
\label{eq:eta}
    \eta = -\frac{1}{2}\frac{r^2}{\mathcal{G}M\rho_\text{m}}\frac{\partial P}{\partial r} = \frac{1}{2}\left(s-\beta+2\right)\left(\frac{h}{r}\right)^2.
\end{equation}

At a sufficiently small size and/or sufficiently high relative velocity, this drag force will act to strongly alter the satellite's orbit, causing the moon to circularize and decay inward \citep{Adachi+76} as
\begin{equation}
\label{eq:gaseccdamptimescale}
    \frac{1}{\tau_e}\bigg|_{\text{gas}} = \frac{\dot{e}}{e}\bigg|_{\text{gas}} = \frac{-1}{\tau_{\text{gas}}}\sqrt{\frac{5}{8}e^2 + \eta^2}
\end{equation}
and
\begin{equation}
\label{eq:gasmigtimescale}
    \frac{1}{\tau_a}\bigg|_{\text{gas}} = \frac{\dot{a}}{a}\bigg|_{\text{gas}} = \frac{2}{\tau_e}\left[\eta+\mathcal{C}_\text{g}e^2\right]
\end{equation}
where $\tau_\text{gas}$ is the characteristic decay timescale defined as
\begin{equation}
\label{eq:taugas}
    \frac{1}{\tau_{\text{gas}}}=\Omega_k \left(C_{\rm{d}}\sqrt{\frac{\pi}{8}}\right)\left(\frac{\Sigma R^2}{m}\right)\left(\frac{h}{r}\right)^{-1},
\end{equation}
and $\mathcal{C}_\text{g}$ is a constant of order unity sheltering the slight dependence on the thermal profile.\footnote{We designate a similar constant $\mathcal{C}_\text{w}$ in the type-I migration expression, done to clean up the bulky equations in \S \ref{sect:formalism}; however, the full dependencies are accounted for in the code running our numerical experiments and can be found in the cited works.} Here $C_{\rm{d}}=0.44$ is the non-dimensional drag coefficient appropriate for high Reynolds number flow, and $\Omega_k = \sqrt{\mathcal{G}M/r^3}$, the keplerian orbital frequency of the satellite. 

Now in the case in which a satellite grows large enough -- such as Io -- the effects of aerodynamic drag become negligible. However, at such large masses, the satellite gravitationally perturbs the gas density distribution. The exact dynamics of this process are complex, but the typical timescale associated with the spiral density wave propagation and the corresponding angular momentum exchange can be approximated in an illustrative manner as \citep{Tanaka&Ward04}
\begin{equation}
\label{eq:tauwave}
    \frac{1}{\tau_{\text{wav}}} = \Omega_k \left(\frac{m}{M}\right)\left(\frac{\Sigma a^2}{M}\right)\left(\frac{h}{r}\right)^{-4}.
\end{equation}
The associated rate of semi-major axis evolution can thus be expressed as
\begin{equation}
\label{eq:typeImig}
    \frac{1}{\tau_a}\bigg|_{\text{typI}} = \frac{\dot{a}}{a}\bigg|_{\text{typI}} \approx \frac{-\mathcal{C}_\text{w}}{\tau_{\text{wav}}}\left(\frac{h}{r}\right)^2.
\end{equation}

Before continuing, let us remark on a few key dependencies in these equations. For a small body on a circularized orbit, the aerodynamic decay timescale of eq. (\ref{eq:gasmigtimescale}) simplifies to $\tau_a|_\text{gas} = -\tau_\text{gas}/(2\eta^2)\propto \Sigma^{-1}(h/r)^{-3}$, with any eccentricity-inducing perturbation correspondingly accelerating the semi-major axis decay, and introducing $\tau_e|_\text{gas}\sim -\tau_\text{gas}/e \propto \Sigma^{-1}(h/r)$. A massive object's type-I migration also depends on these disk properties, with a similar relation to the surface density, but markedly, a distinct relation to the aspect ratio: $\tau_a|_\text{typI}\propto \Sigma^{-1}(h/r)^2$. The key point here is that both $\tau_e|_\text{gas}$ and $\tau_a|_\text{gas}$ scale inversely with $\Sigma$, similar to $\tau_a|_\text{typI}$, yet both maintain a nontrivial, separate dependence on $h/r$ that differentiates them from the type-I regime.

\begin{figure}[h!]
    \centering
    \includegraphics[width=1\linewidth]{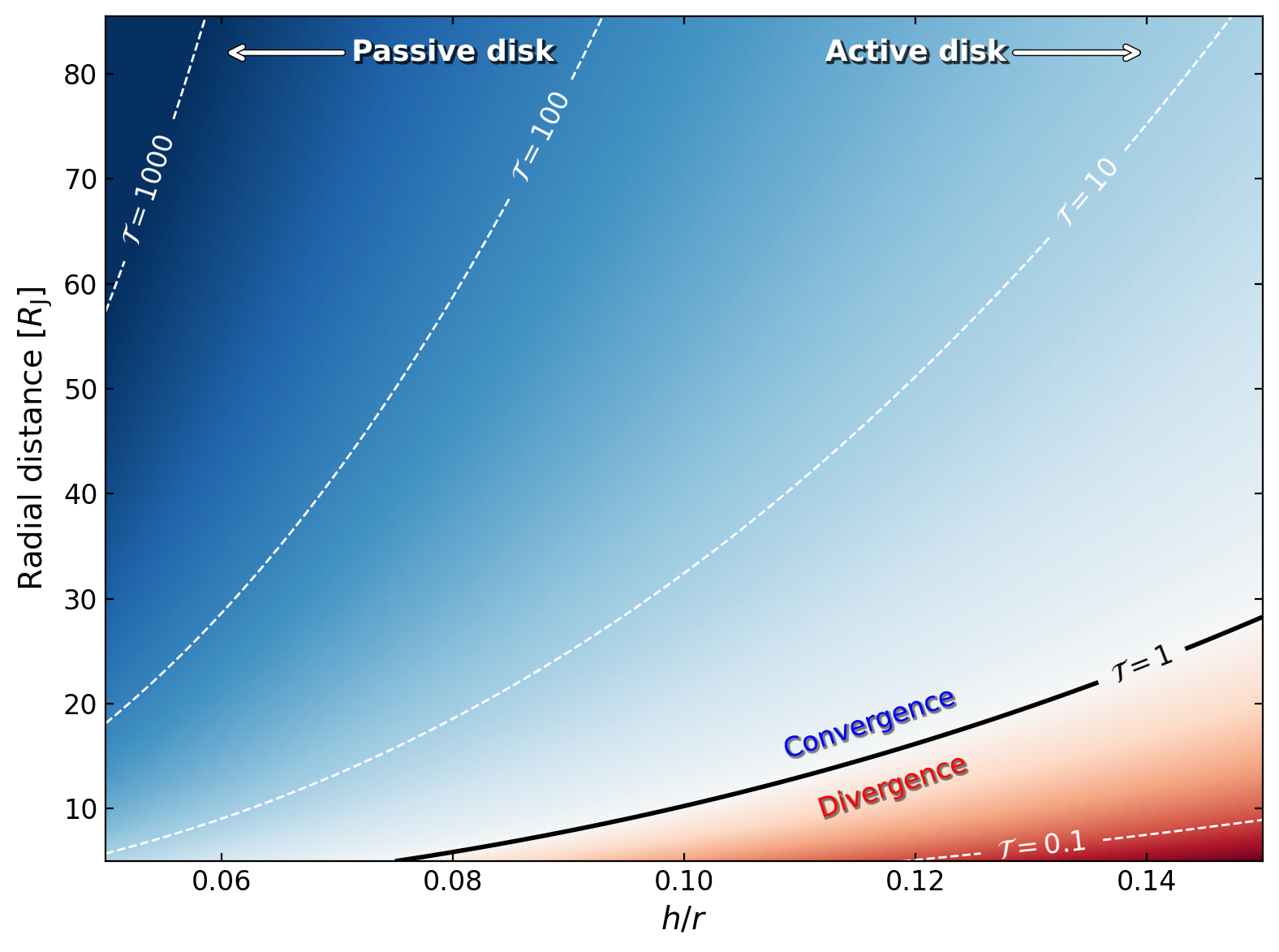}
    \caption{The ratio, $\mathcal{T}$, plotted for various values of $h/r=0.05-0.15$ throughout the approximate location of Amalthea-Io's formation and inner disk edge. Blue corresponds to $\mathcal{T}>1$, areas in which the outer Io catches up to the slower-evolving Amalthea due to the quicker decay from type-I migration. Only in the inner regions of a disk with a high aspect ratio would Amalthea's aerodynamic drag result in a diverging evolution between the two satellites -- shown in red.}
    \label{fig:TRatio}
\end{figure}

In Figure \ref{fig:TRatio}, we have plotted the ratio, $\mathcal{T}$, of the semi-major axis evolution timescales imposed by the aerodynamic drag of a circularized Amalthea, and the type-I migration of Io, which has an appreciable dependence on the disk properties as 
\begin{equation}
\label{eq:MigRatio}
    \mathcal{T} = \frac{\tau_a|_\text{gas}}{\tau_a|_\text{typI}} \propto \frac{\tau_\text{gas}}{\tau_\text{wav}}\left(\frac{h}{r}\right)^2\propto \left(\frac{h}{r}\right)^{-1}.
\end{equation}
The anti-correlation to the aspect ratio (and notable cancellation of $\Sigma$)\footnote{Because the $h/r$ 
 and $\Sigma$ relation are for the two separate bodies, there technically remains a slight $s\ \&\ \beta$ dependence that we are omitting for clarity. Upon cancellation, these dependencies are of order unity. Nonetheless, we do not neglect these terms for our numerical experiments, and they are included in all runs of the N-body simulations below.} illuminates the strong dependence of these satellite dynamics on the thermal profile of the disk. Convergent migration between the small inner body, Amalthea, and large exterior perturber, Io, becomes a necessary occurrence in any reliable disk model, and still occurs more rapidly in a cold, concentrated midplane -- encapsulated by the larger value for $\mathcal{T}$. We also see in Figure \ref{fig:TRatio} that the orbits diverge as Amalthea approaches the disk's inner edge ($\mathcal{T}<1$); however, we note alongside this that an aspect ratio of $h/r\gtrsim 0.12$ so close to the host planet likely becomes physically unrealistic, as this would imply exceptionally high disk temperatures. 

 As the two bodies inevitably converge, the larger Io will capture the small Amalthea into an interior first-order resonance of period ratio, $k':k$, with $k=k'-1$ as a positive integer. This facilitates their co-evolution, effectively shifting Amalthea's rate of decay from the characteristic timescale of aerodynamic drag in eq. (\ref{eq:gasmigtimescale}) to that governed by Io's type-I migration in eq. (\ref{eq:typeImig}), and thereby quickening the small moon's migration through the gaseous medium. The stronger semi-major axis decay is necessarily accompanied by an excitement of its eccentricity, which then as illustrated in equations (\ref{eq:gaseccdamptimescale})--(\ref{eq:taugas}) \textit{enhances} the vigor of the aerodynamic drag. With these qualitative expectations delineated, we will now numerically investigate these competing processes to examine the full, dynamical mechanism of delivering Amalthea to the inner disk via Io's resonant transport.

\section{Numerical Experiments}
\label{sect:NumExp}

We conducted a suite of 20,000 simulations to explore the impact of varying the disk's structure on successful resonant transport of Amalthea to the inner Jovian system. In agreement with the discussion presented in the previous section, we found that reasonable variations in $\Sigma_0$ and $s$ only modulate the \textit{timescale} of overall evolution and not its qualitative character.\footnote{This breaks down if $\Sigma$ is tuned high enough, as this begins to substantially alter the capture probability of Amalthea into resonanace \citep{Batygin&Petit23}. However, such high values for $\Sigma$ are unrealistic for models of the Jovian nebula.} For this reason, we fix our disk profile bearing resemblance to those found throughout the literature, in which we have $\Sigma_0 = 4000$ g/cm$^2$ at $r_0= 0.1$ $R_\text{H}$ and $s=1.25$. The aspect ratio is varied between $h/r=0.05$--0.15, and kept constant for each run ($\beta=0$). Io is initialized on a circular, planar orbit at a distance approximating the satellitesimal formation zone described in \S \ref{sect:Just} ($a'_i = r_0$). The small satellite Amalthea is then placed on a near-circular orbit in the vicinity of the inner edge of this zone, in which we vary its exact distance between $a_i=0.75$--$0.95\ a'_i$. We rely on the N-body numerical integration library \texttt{Rebound} \citep{rebound}, with the integrator \texttt{mercurius} \citep{reboundmercurius}.\footnote{\hyperlink{https://github.com/hannorein/rebound}{https://github.com/hannorein/rebound}} Non-gravitational effects, such as type-I migration and gas drag, are implemented via the supplemental library, \texttt{Reboundx} \citep{reboundx}.\footnote{\hyperlink{https://github.com/hannorein/rebound}{https://github.com/dtamayo/reboundx}} The type-I migration implementation follows that of \citet{Kajtazi+23}, based on \citet{Cresswell&Nelson08} and \citet{Pichierri+18}, in agreement with the above eq. (\ref{eq:tauwave}) and (\ref{eq:typeImig}). The effects of aerodynamic drag are implemented following eq. (\ref{eq:gaseccdamptimescale})--(\ref{eq:taugas}).

With Io's migration, gravitational dynamics, and Amalthea's gas drag modeled in a self-consistent manner, we evolve the system until one of three occurrences: (1) Amalthea reaches the inner disk edge, (2) Amalthea and Io collide, or (3) Amalthea is scattered outward and passed by Io. Outcomes (2) and (3) indicate ``failure" of the resonant transport mechanism, whereas (1) indicates ``success," with Amalthea safely reaching its current orbital neighborhood. 

\begin{figure*}
    \includegraphics[width=\textwidth]{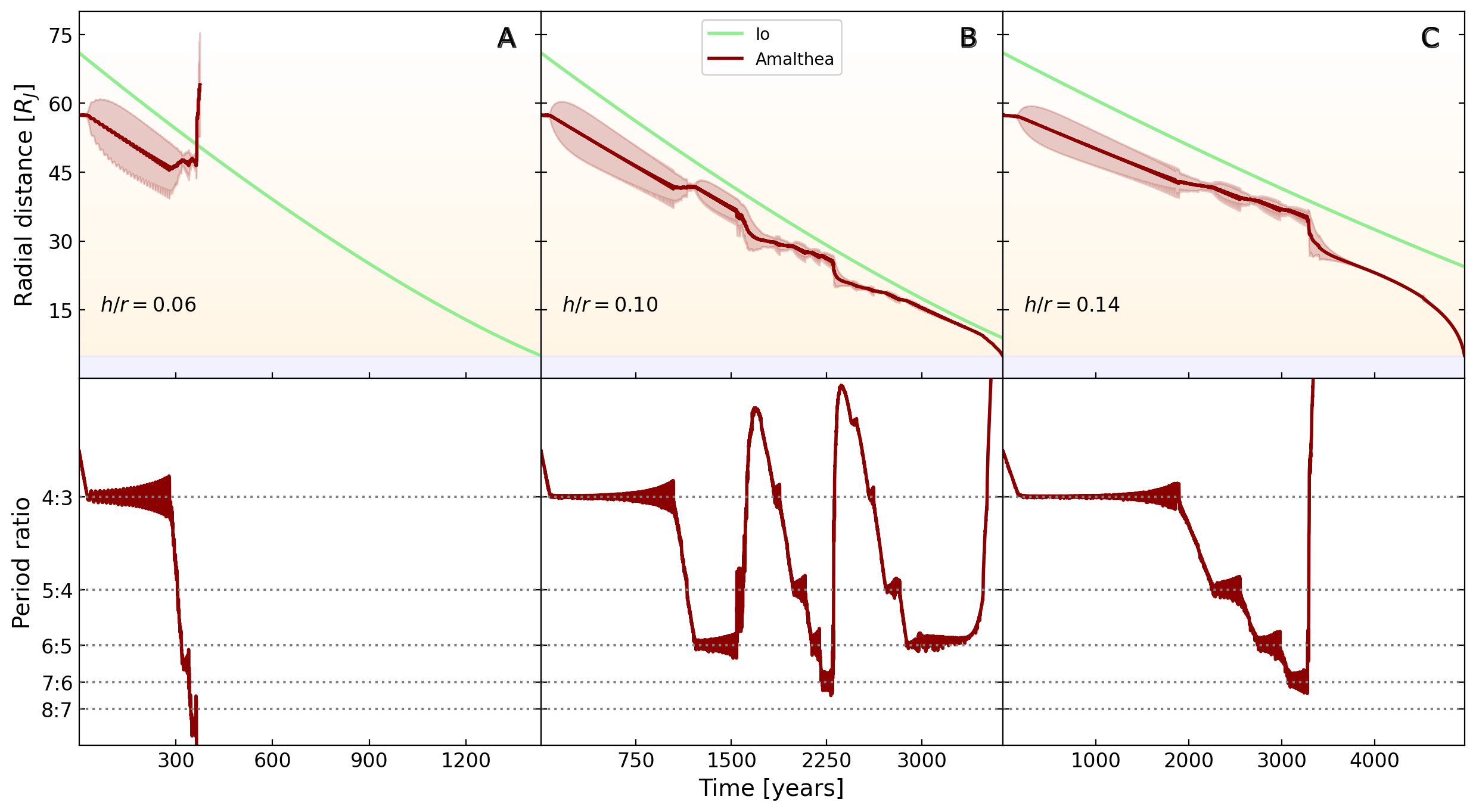}
    \caption{Three simulations following the scenario described in the text with the same parameters except the noted variation in the disk aspect ratio, $h/r$. Io and Amalthea form in a zone of satellite building blocks near $r_0\sim0.1\ R_\text{H}$. As Io (light green) grows large enough to type-I migrate, it captures the interior satellitesimal, Amalthea (dark red), into resonance. Top and bottom panels display the time series of the radial distance and Io-Amalthea period ratio, respectively. The lightly shaded red envelope around the darker semi-major axis line of Amalthea in the top panels traces the difference between Amalthea’s perijove, $r_p$, and apojove, $r_a$, essentially characterizing the moon’s eccentricity ($e\sim0.1$ in the initial 4:3 resonance). This eccentricity grows in resonance while being dampened by aerodynamic drag. A light blue band serves to mark the magnetospheric cavity, beginning at the disk's truncation radius, $R_\text{t}=5\ R_\text{J}$. The resonant configuration is broken as a consequence of overstability. These overstable librations repeatedly hinder resonant transport, often leading to the outward scattering of Amalthea (as in \textbf{A}) with the disk aspect ratio, $h/r$, playing a critical role in determining the likelihood of this failure.}
    \label{fig:MigPanels}
\end{figure*}

Figure \ref{fig:MigPanels} displays three exemplary runs of these simulations, each with Amalthea slightly interior of the 4:3 resonance to start, and as such, with inputs differentiated only by the disk's $h/r$ value of 0.6, 0.10, 0.14 (panel A, B, and C). Notably, in each of these simulations, Amalthea is immediately captured into the 4:3 resonance due to the convergent migration (as indicated by the large $\mathcal{T}$ from Figure \ref{fig:TRatio}). Subsequently, Amalthea's eccentricity is pumped up to an equilibrium value, where it then begins to oscillate until the amplitude of this oscillation grows so large that the resonance breaks and convergent migration between the two satellites resumes. After this initial break, in the runs displayed in Figure \ref{fig:MigPanels}B and \ref{fig:MigPanels}C, Amalthea is quickly recaptured into a closer resonance (where it repeats the cycle again), and Io goes on to successfully deliver the small moon to the magnetospheric cavity. In Figure \ref{fig:MigPanels}A's run, however, upon the initial resonance breaking, Io is unable to successfully recapture the small moon back into resonance; the orbits quickly converge, and then amidst this chaotic sea of compact orbits, Amalthea is scattered outward back into the satellite formation region beyond $r > 0.1$ $R_\text{H}$. 

\begin{figure}
    \centering
    \includegraphics[width=1\linewidth]{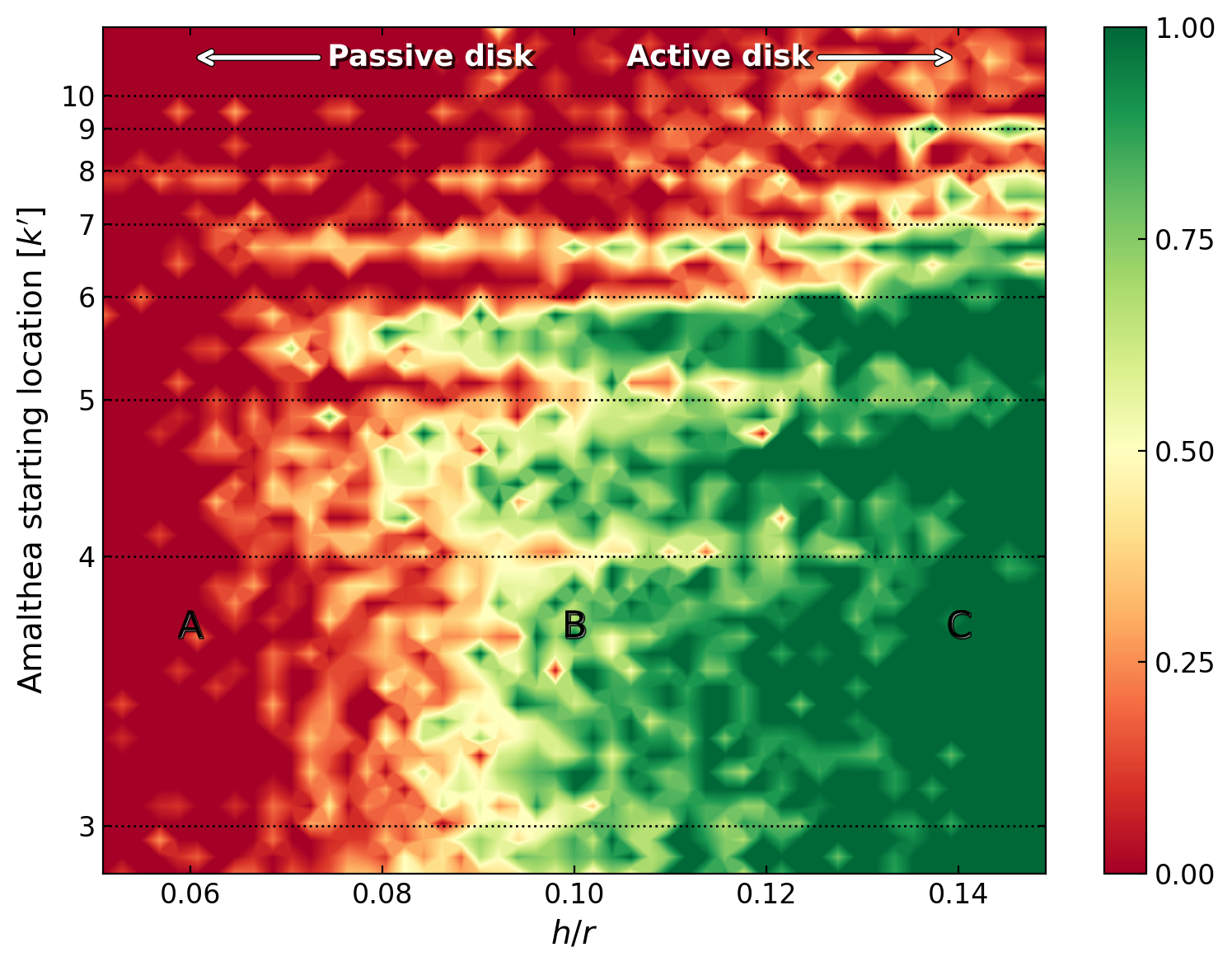}
    \caption{A grid of outcomes for 20,000 simulations, varying the initial starting location and aspect ratio as described in the text. Color corresponds to the success rate of Io's delivery of Amalthea to the inner disk edge. Points A, B, and C correspond to examples shown in Figures \ref{fig:MigPanels} and \ref{fig:43Pan}.}
    \label{fig:hMap}
\end{figure}

Figure \ref{fig:hMap} displays the grid of outcomes for all 20,000 of our simulations, with color indicating the success rate of the delivery. The exemplary runs from Figure \ref{fig:MigPanels} are indicated on the map by their labels, A, B, and C. We see in these results a distinct influence of the disk structure on Amalthea's long-range resonant transport. As $h/r$ is lowered -- i.e., a passive, cooler disk profile -- the likelihood of successful delivery of the small body to the magnetospheric cavity becomes improbable. While Figure \ref{fig:hMap} makes it clear that the aspect ratio plays a critical role in the numerical experiments, let us now move to analytical grounds to explore the fundamental link between the viability of long-range resonant transport and the disk scale height. 

\section{Perturbation Theory}
\label{sect:formalism}

The reason for the repeated failures seen in our numerical experiments, is that no matter the structure of the disk, the evolving resonance between Io and Amalthea is overstable. While resonance locking facilitated by convergent migration has been understood for over half a century \citep{Goldreich65,Henrard&Lemaitre83,Neishtadt84}, studies examining \textit{overstable librations} have only entered the literature in the last decade -- albeit for angular-momentum-conserving dissipative forces (such as tides and type-I torques in \citealt{Goldreich&Schlichting14,Deck&Batygin15,Xu+18}). In this regard, the Io-Amalthea resonance is quantitatively distinct, since aerodynamic dissipation drives nonlinear eccentricity damping. 

Qualitatively, overstability unfolds when a less massive satellite resides interior to a larger, converging body. As the system enters a nominal resonance, the interplay between the orbital migration and eccentricity damping drives the system to a global fixed point, characterized by an equilibrium eccentricity, $e_0$ \citep{Goldreich&Schlichting14}. The Tisserand relation dictates that the interior body's eccentricity and semi-major axis variations are anti-correlated, so a stronger semi-major axis damping at times of high eccentricity leads to a gradual amplification of the libration amplitude around the resonance. If these oscillations grow too large, the small body enters the domain of overlap with a neighboring mean-motion resonance, inducing chaotic evolution \citep{Wisdom80,Hadden&Lithwick18}. Let us now quantify these dynamics from perturbative grounds.

\subsection{Resonant Hamiltonian}

As in our numerical experiments, we approximate a perturber (Io) on a circular orbit exterior to a smaller body of negligible mass (Amalthea). The governing Hamiltonian in the restricted three-body problem, $\mathcal{H}$, for a first-order interior resonance takes the summed form of its keplerian orbit and resonance perturbation,
\begin{equation}
\label{eq:HammyKep}
  \begin{aligned}
    \mathcal{H} & = \mathcal{H}_{\text{k}} + \mathcal{H}_{\text{r}}\\
      & = \frac{-\mathcal{G}M}{2a} - \frac{\mathcal{G}m'}{a'}fe\cos{\left(k'n't - k\lambda - \varpi\right)},
  \end{aligned} 
\end{equation}
where $\mathcal{G}$ is the gravitational constant, $M$ the mass of the central body (Jupiter), and $a$ the semi-major axis of the satellite. Primed variables refer to the exterior perturber, Io (unprimed, the massless inner-body, Amalthea). In the perturbative term of the Hamiltonian we have $e$, the eccentricity, and the coefficient, $f$, as a negative constant of order unity.\footnote{$f$ being dependent on the semimajor axis ratio, $a/a'$, with a negative value for an external perturber, $f\approx -0.8\ k'$, for $k'\geq 2$.} The resonant argument within the cosine has time, $t$, the mean motion, $n$, the mean longitude, $\lambda$, and the longitude of perijove, $\varpi$.

We adopt the canonical approach and utilize the Poincaré action-angle variables,

\begin{equation}
\label{eq:Poincaré}
  \begin{aligned}
  \Lambda&=\sqrt{\mathcal{G}Ma} & \lambda &= \mathcal{M} + \varpi\\
    \Gamma &= \Lambda\left(1-\sqrt{1-e^2}\right) & \gamma &= -\varpi
  \end{aligned} 
\end{equation}
where $\lambda$ and $\gamma$ are now the coordinates corresponding to their respective conjugate momenta, $\Lambda$ and $\Gamma$. From eq. (\ref{eq:HammyKep}), the keplerian term now becomes, $\mathcal{H}_{\text{k}} = -(\mathcal{G}M/\Lambda)^2/2$.

We are specifically interested in the evolution of dynamics near nominal resonance (subscript nought), $n'/n_0 = k/k'$. As such, the first nominal action becomes $\Lambda_0 = \sqrt{\mathcal{G}Ma'}\left(k/k'\right)^{1/3}$, and the second action at the relatively small eccentricities of concern here is approximated as $\Gamma\approx\Lambda_0(e^2/2)$. Upon expansion of $\mathcal{H}_{\text{k}}$ in the vicinity of interest to second order in $\delta\Lambda = \Lambda-\Lambda_0$, we can rewrite the Hamiltonian with the Poincaré variables as, 
\begin{equation}
\label{eq:HamPoincaré}
  \begin{aligned}
    \mathcal{H}_p = &\ \frac{(\mathcal{G}M)^2}{\Lambda_0^3}\delta\Lambda - \frac{3(\mathcal{G}M)^2}{2\Lambda_0^4}\left(\delta\Lambda\right)^2\\
    &\ - \mathcal{A}\sqrt{2\Gamma}\cos{(k'n't-k\lambda+\gamma)} + \rm{T},
  \end{aligned}
\end{equation}
where we defined $\mathcal{A}= \mathcal{G}^2Mm'f\left(k/k'\right)^{2/3}/\Lambda_0^{5/2}$, and extended phase space with an action, T, conjugate to time, making the Hamiltonian formally autonomous. 

Next, we perform a canonical transformation of variables with a type-2 generating function, 
\begin{equation}
\label{eq:GenFunc}
  \begin{aligned}
    \mathcal{F}_2 = (k'n't - k\lambda + \gamma)\Phi + \lambda\Psi + t\Xi,
  \end{aligned} 
\end{equation}
which now yields new action-angle variables of 
\begin{equation}
\label{eq:Actions}
  \begin{aligned}
\Phi&=\Gamma & \phi &= k'n't - k\lambda + \gamma\\
\Psi &= \Lambda + k\Phi & \psi &= \lambda\\
\Xi &= \text{T} - k'n'\Phi & \xi &= t. 
  \end{aligned} 
\end{equation}
With this transformation and a bit of algebraic manipulation, the Hamiltonian from eq. (\ref{eq:HammyKep}) takes on the simple functional form of
\begin{equation}
\label{eq:HammyFinal}
    \mathcal{H}' = 2\mathcal{B}(\Psi-\Lambda_0)\Phi - \mathcal{B} k\Phi^2 - \mathcal{A}\sqrt{2\Phi}\cos(\phi),
\end{equation}
with $\mathcal{B}=3(\mathcal{G}M)^2k/(2\Lambda_0^4)$. Lastly, we can rescale $\mathcal{H}'$ by $\mathcal{B}\Lambda_0^2$ (just equivalent to $3k\mathcal{H}_\text{k}$), eliciting a dimensionless
\begin{equation}
\label{eq:HammyFinalRed}
    \tilde{\mathcal{H}} = 2(\tilde{\Psi}-1)\tilde{\Phi} - k\tilde{\Phi}^2 - \chi\sqrt{2\tilde{\Phi}}\cos(\phi),
\end{equation}
with the positive constant, $\chi=\frac{\mathcal{A}}{\mathcal{B}\Lambda_0^{3/2}} = \frac{-2}{3}\frac{m'}{M}\frac{f}{k}\left(\frac{k}{k'}\right)^{2/3}$ and each action scaled as $\tilde{\mathcal{S}}=\mathcal{S}/\Lambda_0$. This casts the simplified Hamiltonian into a form equivalent to the second fundamental model for resonance \citep{Henrard&Lemaitre83}. Moreover, the term ($\tilde{\Psi}\rm{-}1$) encapsulates the deviation from nominal resonance, and thus given a value of $\tilde{\Psi}$, there exists a corresponding value of the action, $\tilde{\Phi}_0$, that is characteristic of zero-amplitude libration about the equilibrium. In practice, the value of $\tilde{\Phi}_0$ is sufficiently high to approximate $\tilde{\mathcal{H}}$ as a pendulum, and the equations of motion are
\begin{equation}
\label{eq:HammyEOMs}
\ddot{\tilde{\Phi}} =\chi\sqrt{2\tilde{\Phi}_0}\cos{(\phi)} \qquad \ddot{\phi} = -\omega_0^2\sin{(\phi)}
\end{equation}
with a maximal libration half-width of 
\begin{equation}
\label{eq:LibWidth}
    \Delta a = \pm\ a_0\left(8k'|\chi|e_0\right)^{1/2}
\end{equation}
and librational frequency, $\omega_0 = (2k|\chi| e_0)^{1/2}$.

\subsection{Overstability}

Because the Hamiltonian is not a function of the angle, $\psi$, the action $\tilde{\Psi}$ is a constant of motion representing the Tisserand constant in the sense that gravitational dynamics do not modulate its evolution. The exact value of $\tilde{\Psi}$ is instead driven by the forces of dissipation induced by the disk alone. 

\subsubsection{Equilibrium eccentricity}

Ultimately, the competing processes of dissipative damping and resonant forcing give rise to an equilibrium eccentricity, $e_0$, which we will analytically derive here. Let's start by rewriting the approximate Tisserand relation in keplerian elements as $\tilde{\Psi} = \sqrt{a/a_0} + ke^2/2$ and thus 
\begin{equation}
\label{eq:dotPsi}
    \dot{\tilde{\Psi}} \approx \frac{1}{2}\left(\frac{\dot{a}}{a}\bigg|_{\text{gas}}-\frac{\dot{a}_0}{a_0}\bigg|_{\text{typI}}\right) + ke^2 \frac{\dot{e}}{e}\bigg|_{\text{gas}}.
\end{equation}
Recall these terms from the discussion in \S \ref{sect:NumExp}. Here, $\dot{e}/e$ and $\dot{a}/a$ correspond to the eccentricity damping and semi-major axis decay imposed by aerodynamic drag (eq. (\ref{eq:gaseccdamptimescale}) and (\ref{eq:gasmigtimescale}), respectively); whereas $\dot{a}_0/a_0$ corresponds to the inverse migrational timescale from type-I migration (eq. (\ref{eq:typeImig})), which in this case, manifests through Io's resonant influence on the smaller body. Since $\dot{\tilde{\Psi}}=0$ at the global fixed point, we can exploit eq. (\ref{eq:dotPsi}) to reveal the equilibrium value, $e_0$. Upon making the slight simplification that $\sqrt{5e^2/8 + \eta^2}\approx \sqrt{5e^2/8}$ inside $\dot{e}/e$ and $\dot{a}/a$, we find a rather unintuitive expression, 
\begin{equation}
\label{eq:eqecc}
\begin{aligned}
    e_0 = \frac{\daleth^{1/3}}{3(k+\mathcal{C}_g)}-\frac{\eta}{\daleth^{1/3}},
\end{aligned}
\end{equation}
with the bulky 
\begin{equation*}
\label{eq:daleth}
\begin{aligned}
    \daleth &= \frac{3}{\sqrt{10}\tau_\text{wav}}\Biggl[9\mathcal{C}_\text{w}\left(\frac{h}{r}\right)^2(k+\mathcal{C}_\text{g})^2\tau_\text{gas}\ +\\
    &\sqrt{3(k+\mathcal{C}_\text{g})^3\left(27\mathcal{C}_\text{w}^2\left(\frac{h}{r}\right)^4 (k+\mathcal{C}_\text{g})\tau_\text{gas}^2+10\eta^3\tau_\text{wav}^2\right)}\Biggr].
\end{aligned}
\end{equation*} 
While eq. (\ref{eq:eqecc}) does agree with the eccentricity in the numerical experiments (as we will display in the figures below), its form is of little use for any qualitative understanding. Let us instead examine a pair of illustrative limits. First, in which $\eta\ll e^2$ (relevant for much of the cooler disk profiles), yielding a much simpler
\begin{equation}
\label{eq:eqeccsmalleta}
\begin{aligned}
    e_0 \approx \left[\frac{\mathcal{C}_\text{w}}{k+\mathcal{C}_\text{g}}\frac{\tau_\text{gas}}{\tau_\text{wav}}\left(\frac{h}{r}\right)^{2}\right]^{1/3}\ \propto\ \left(\frac{h}{r}\right)^{-1/3};
\end{aligned}
\end{equation}
and second, in which $\eta\rightarrow\mathcal{C}_\text{g}e^2$ (relevant for the warmer disk profiles), still similarly yielding,
\begin{equation}
\label{eq:eqeccetaequal}
\begin{aligned}
    e_0 \approx \left[\frac{\mathcal{C}_\text{w}}{k+2 \mathcal{C}_\text{g}}\frac{\tau_\text{gas}}{\tau_\text{wav}}\left(\frac{h}{r}\right)^{2}\right]^{1/3}\ \propto\ \left(\frac{h}{r}\right)^{-1/3},
\end{aligned}
\end{equation}
where the slightly larger term in the denominator underscores the increasing influence of the eccentricity damping with $\eta$.

Although not precise, eq. (\ref{eq:eqeccsmalleta}) and (\ref{eq:eqeccetaequal}) better highlight the key relationship in this regime of discrepant satellite-disk dynamics that stands markedly in contrast from that of the oft-studied regime in which both satellites are massive enough for type-I damping. In the mutual type-I comparison the eccentricity excitation remains comparable to the disk aspect ratio, evolving as $e_0\sim h/r$ \citep{Goldreich&Schlichting14,Pichierri+18,Batygin&Petit23}. But now, for a small body subject to aerodynamic drag, the $e_0$ dependence on $h/r$ is reciprocal.

\begin{figure*}
\includegraphics[width=\textwidth]{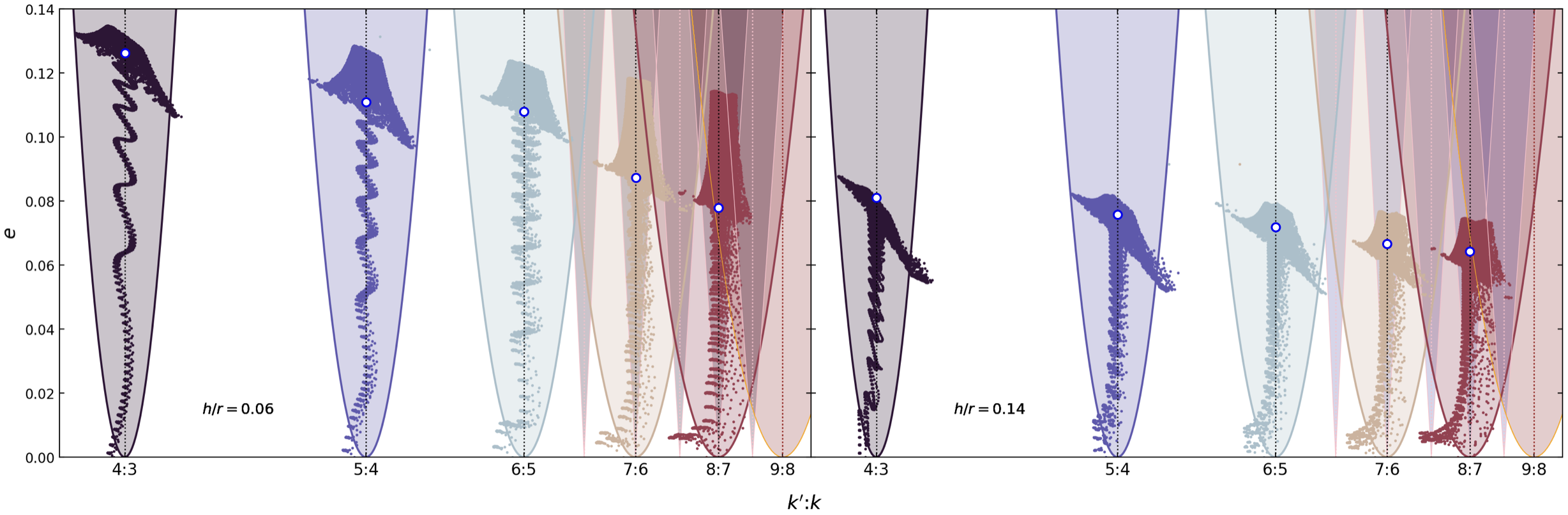}
    \caption{Here we show the resonant encounters for $k'=4$--$9$, with $h/r=0.06$ on the left, and $h/r=0.14$ on the right. Colored regions represent the theoretical width of first-order mean-motion resonances from eq. (\ref{eq:LibWidth}), with the pink thinner widths being those of the second-order resonances at more compact orbits. Vertical dotted lines indicate the nominal resonant semi-major axis, $a_0$, and the blue-bordered white dots represent the value of the analytically-derived equilibrium eccentricity from eq. (\ref{eq:eqecc}). Note the propensity for overlap at the high eccentricities of more compact orbits, an area increasingly encountered in a regime of low $h/r$.}
    \label{fig:ResCross}
\end{figure*}

A profound consequence of this relation arises in consideration of the neighboring resonances, in which the chaotic zone of resonance overlap \textit{widens} with the excitement of a satellite's eccentricity \citep{Wisdom80,Hadden&Lithwick18}. Therefore, the reciprocal relation between $h/r$ and $e_0$ means that when the Io-Amalthea resonance breaks, if the disk aspect ratio is low, the small body is prone to be thrust into a chaotic regime with a high $e$. To illustrate this relation, in Figure \ref{fig:ResCross} we show $e$ vs. $a$-space for characteristic simulations (at $h/r=0.06$ and 0.14) of Amalthea evolving in first-order resonance. We overplot the first, and neighboring second-order\footnote{We modelled our Hamiltonian around an approximate mapping of the first-order interior resonant dynamics; however, we can use eq. (\ref{eq:LibWidth}) to loosely depict the second-order resonances for heuristic purposes.} resonance widths using eq. (\ref{eq:LibWidth}). In the vicinity of more compact orbits (high $k'$), the amplified eccentricity thrusts Amalthea into the chaotic sea, where it becomes increasingly likely to be scattered back outward as the resonance cannot be restablished. 

\subsubsection{Growth of librations}

As we see in Figure \ref{fig:MigPanels} and \ref{fig:ResCross}, it is not merely the magnitude of the eccentricity that is pumped up with the lower $h/r$, but also the growth rate of the amplitude of librations about the nominal resonance is accelerated. Again, recalling that our Hamiltonian resembles that of a pendulum, we will now characterize the overstability of the librations analytically.

Let us define an energy bandwidth, $\Delta\mathcal{H} = \mathcal{H}_{\text{h}} - \mathcal{H}_{\text{e}}$, as the difference between the maximal libration state,  $\mathcal{H}_{\text{h}}$ (corresponding to the hyperbolic fixed-point), and the minimal libration state, $\mathcal{H}_{\text{e}}$ (corresponding to the elliptic fixed-point). Suppose the system has just entered a resonant state, and thus the value of the Hamiltonian sits safely in the libration regime close to $\mathcal{H}_{\text{e}}$. A measure of the amplitude of libration at this moment is then $A = (\mathcal{H}-\mathcal{H}_{\text{h}})/\Delta\mathcal{H}$, since passing beyond $\mathcal{H}_{\text{h}}$ would send the body into the circulating regime. At the elliptic and hyperbolic fixed-points, $\tilde{\Phi} = \tilde{\Phi}_0$ for both, and $\phi = 0$ and $\phi = \pi$, respectively, yielding
\begin{equation}
\label{eq:amp}
\begin{aligned}
    A = &\frac{1}{2\chi\sqrt{2\tilde{\Phi}_0}}\left(2(\tilde{\Psi}-1)(\tilde{\Phi}-\tilde{\Phi}_0)-k(\tilde{\Phi}^2-\tilde{\Phi}_0^2)\right)\\
    &+ \sin^2(\phi/2).
\end{aligned}
\end{equation}

\begin{figure*}
    \includegraphics[width=\textwidth]{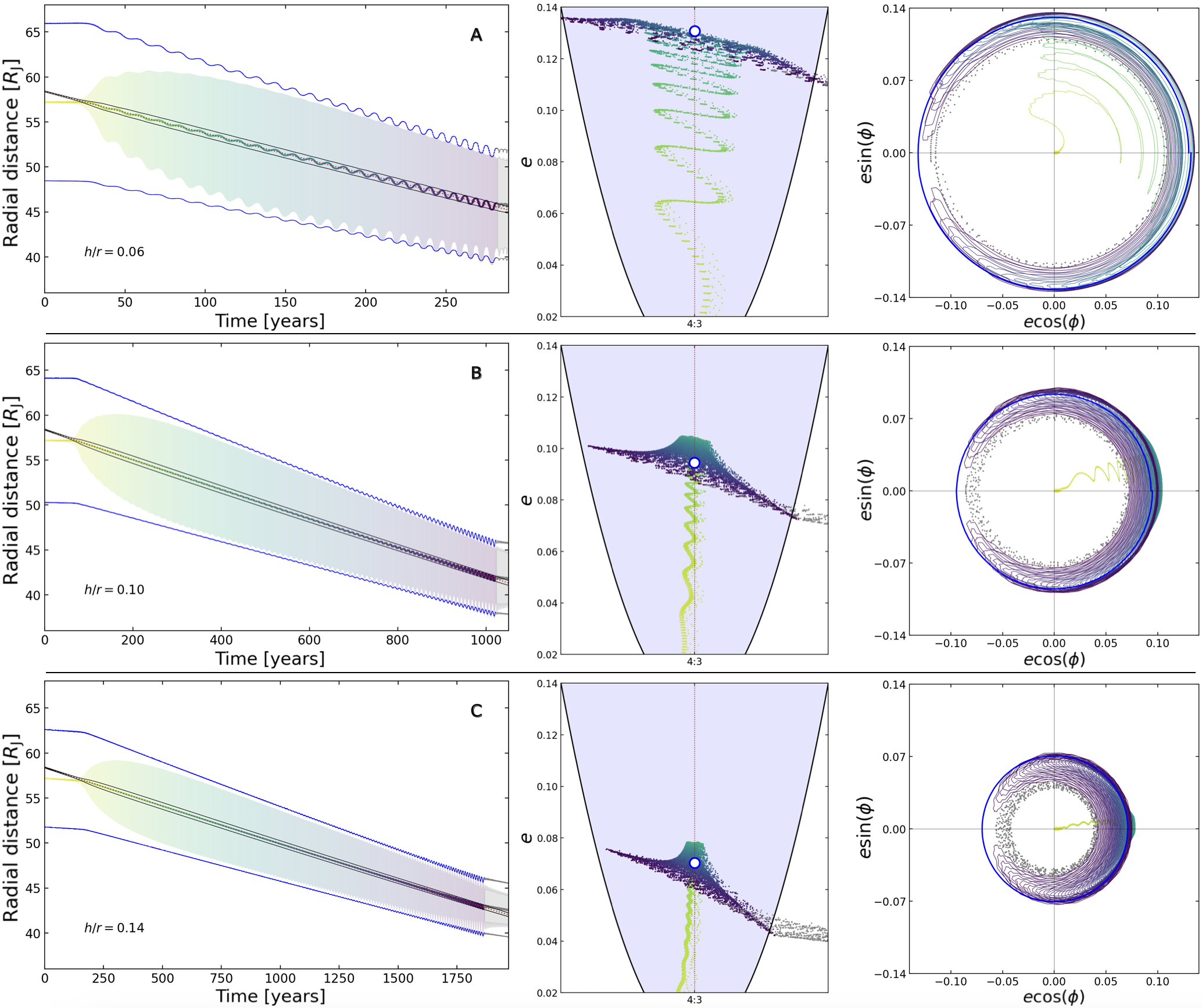}
    \caption{The same simulations from Figure \ref{fig:MigPanels}, now displaying the resonant dynamics of the initial 4:3 only. In the top plot of each panel, the semi-major axis evolution is displayed with color corresponding to the time labeled on the x-axis. The shaded region again represents the orbital excursion between perijove and apojove, illustrating the evolution of Amalthea's eccentricity. Blue lines indicate the analytically-derived equilibrium peri-/apojove distances that correspond to the equilibrium eccentricity, $e_0$, from eq. (\ref{eq:eqecc}). Black lines mark the boundaries of the resonance width, $\Delta a$ from eq. (\ref{eq:LibWidth}). The bottom left panel then plots $e$ vs $a/a'$, with $\Delta a$ shaded and the blue-white dot corresponding to $e_0$ from eq. (\ref{eq:eqecc}). The bottom right panel projects the evolution into the phase-space diagram, tracing the evolution of Amalthea's eccentricity vector in a reference frame rotating with the resonant angle, where the blue circle represents $e_0$. }
    \label{fig:43Pan}
\end{figure*}

In a purely Hamiltonian system (where $\dot{\mathcal{H}}=0$), $A$ would remain constant. But instead, amidst the gaseous disk, like $\tilde{\Psi}$, the evolution of the amplitude, $\dot{A}$, is entirely driven by dissipative effects which act independently of the angle, $\phi$. Ergo, with the sine term in eq. (\ref{eq:amp}) dropping out, we derive, 
\begin{equation}
\label{eq:ampdot}
\begin{aligned}
    \dot{A} = \frac{1}{\chi\sqrt{2\tilde{\Phi}_0}}(\dot{\tilde{\Psi}}-k\dot{\tilde{\Phi}})(\tilde{\Phi}-\tilde{\Phi}_0).
\end{aligned}
\end{equation}
Although this expression is well-determined given the defined variables above, its final form is cumbersome. Thus, to fix ideas in the spirit of linear stability analysis, we may consider the leading order perturbation, $\tilde{\Phi} \approx \tilde{\Phi}_0(1+\epsilon\sin(\omega_0 t))$, where $\epsilon\ll1$. Upon substitution and expansion of this term, we then have
\begin{equation}
\label{eq:ampdotexp}
\begin{aligned}
    \dot{A} = &\frac{\sqrt{\tilde{\Phi}_0}}{2\chi\sqrt{2}}\left(\frac{\mathcal{C}_\text{w}\left(\frac{h}{r}\right)^2}{\tau_\text{wav}} - \frac{\sqrt{5\tilde{\Phi}_0}(\eta+2\mathcal{C}_\text{g}\tilde{\Phi}_0)}{\sqrt{2}\tau_\text{gas}}\right)\epsilon\sin(\omega_0 t)\\
    &-\frac{\sqrt{5}\tilde{\Phi}_0(\eta+6\mathcal{C}_\text{g}\tilde{\Phi}_0)}{8\chi\tau_\text{gas}}\epsilon^2\sin^2(\omega_0 t) + \mathcal{O}(\epsilon)^3.
\end{aligned}
\end{equation}
To compute the secular drift in $A$, we average $\dot{A}$ over a single libration cycle, $P=2\pi/\omega_0$, which simplifies greatly due to symmetry of the sine term, and upon some rearrangement we attain
\begin{equation}
\label{eq:ampdotavg}
\begin{aligned}
\langle\dot{A}\rangle &= \frac{1}{P}\int_0^P \dot{A}\ dt =\frac{-\epsilon^2\sqrt{5}e_0^2}{32\chi\tau_\text{gas}}(\eta+3\mathcal{C}_\text{g}e_0^2)>0.
\end{aligned}
\end{equation}

This expression is positive definite, proving the growth of the librations as Amalthea enters resonance.\footnote{Recall that $\chi\propto f\propto-k'$, therefore $-\chi>0$.} Moreover, calling upon the simplified relation of $e_0$ and $h/r$ from eq. (\ref{eq:eqeccsmalleta}) and (\ref{eq:eqeccetaequal}), we can see that the first term goes approximately as $\propto (h/r)^{1/3}$, but the second, dominant term evolves $\propto(h/r)^{-7/3}$; elucidating both the amplified magnitude and quickening of the librations when $h/r$ is lowered. 

To link these analytic estimates to the numerical simulations, let us visualize more clearly the impact of $h/r$ on $e_0$ and $\langle\dot{A}\rangle$. The three panels in Figure \ref{fig:43Pan} correspond again to the A, B, and C simulations labeled in Figure \ref{fig:hMap}, now zoomed in to the specific 4:3 resonance from which Amalthea broke out. Most evident from these relations is the increased eccentricity, along with the amplified libration due to the concentration of gas in the midplane as $h/r$ is varied from 0.06 to 0.14. 

\section{Discussion}
\label{sect:Discussion}

In this work, we have considered the dynamical origins of Amalthea, while outlining a mechanism by which it could be delivered to its modern-day orbital neighborhood during the disk-bearing epoch of Jupiter's evolution -- one that naturally accounts for the anomalous features of the small icy moon, the properties of the circumjovian disk, and the constraints set by the neighboring Galileans. In our proposed picture, Io and Amalthea experience starkly different rates of gas-driven evolution within the disk, resulting in convergent migration that precipitates resonant capture. The resonant dynamics are complicated by the damping of Amalthea's eccentricity from the surrounding gas, in which overstable librations around the equilibrium eccentricity lead to breaking of the resonant dynamics. Depending on the disk aspect ratio, $h/r$, these overstable librations thus lead to either a cycle of repeated capture into mean motion resonance, followed by overstable escape, or to an overall failure in delivering Amalthea to the inner cavity of Jupiter's disk. Let us review the implications of our results first for the circumjovian disk, then briefly address the remaining natural satellites of Jupiter, and finally remark on the translation of these results to comparable astrophysical environments both in our solar system and others.  

\subsection{Constraints on the circumjovian disk}

As we have discussed, theoretical models for Jovian satellite formation are primarily constrained by the known properties and orbits of the Galilean moons. Here we have expanded upon these theories to account for the formation of the inner moons, revealing new constraints that do not emerge from modeling the Galilean satellites alone. The figures in \S \ref{sect:NumExp} and \S \ref{sect:formalism} clearly show the stark dependence on $h/r$. In particular, values lower than $h/r \lesssim 0.08$ seem to preclude the survival of Amalthea in the inner Jovian system. This result is significant, since the aspect ratio is intimately linked to the thermal structure of the disk itself as discussed in \S \ref{sect:Just}. We have adopted values for the parameter space used in our simulations based on noted models of satellite formation in the literature. However, we do acknowledge that the specific result of $h/r \lesssim 0.08$ is not \textit{entirely} model-independent. We have more or less taken on the assumptions that the formation of a low-density, icy body \textit{must} occur in the outer regions of the disk, and that the Galilean satellites coalesced entirely (or almost entirely) at a roughly similar distance. This latter assumption is not necessarily implied in all models; particularly, those invoking pebble accretion as the primary mechanism for satellite formation. 

\subsection{Thebe, Adrastea, and Metis}

We have focused solely on the largest member of the Amalthea group, but the three remaining regular satellites -- Thebe, Adrastea, and Metis -- possess intriguing features in themselves. However, less can be definitively said about them since each have an even thinner data set than that of Amalthea. No measurement of their mass has been attempted, so it remains to be seen if they match the low density. One could easily extrapolate and assume similar density and surface compositional profiles for that of Thebe, and perhaps even for the inner-ring moons of Metis and Adrastea. This would entail that these small bodies were concurrently captured into resonance during Io's inward migration, and would reflect similar dynamical histories as that explored here. Nonetheless, it is this exact line of thinking that led previous research astray in assuming that Amalthea would align with the compositional gradient of its larger neighbors.

\citet{Takato+04} did take spectra of Thebe's trailing side over a narrower range of wavelengths than that of Amalthea, finding a similar spectral slope between the two in this range. \citet{Cuk+23} have further surmised that in order for Thebe to remain immune from sesquinary catastrophe, it must either have significant internal strength, or be unexpectedly much more dense than Amalthea. Interestingly, \citet{Tiscareno+13} and \citet{Hedman15} determined that Metis and Adrastea so too must have a higher internal strength to remain intact against tides if they indeed possess a similar density to that of Amalthea. While Thebe's position between Io and Amalthea indicates it would be included in the resonant transport mechanism, it is quite possible that the smaller Metis and Adrastea -- separated from each other by only $\Delta a\sim$ 1000 km -- possess distinct dynamical histories from that of their larger neighbors. These are scenarios that are beyond the scope of this work, but we remark here that these bodies are far too small to notably alter the orbits of either Amalthea or Io in the numerical experiments above. 

\subsection{Other Systems}

By exploring this realistic scenario for Amalthea, we have illuminated more general implications for small bodies in astrophysical disks. The results and perturbative model presented above reveal that the thermodynamics and structure of the disk can readily impede upon the existence of small bodies on short-period orbits in planetary/satellite systems. More specifically, concentration of gas in the midplane adversely affects the long-range transport of small bodies. As a disk cools and its vertical structure consolidates from a lack of thermal support, it becomes less feasible for a large, migrating body to transport small, interior satellites. Instead, the smaller neighbors are forced into a chaotic regime, and either forced outward to an exterior orbit, or -- if there were perhaps multiple planet-/satellitesimals -- scattered into collisional orbits.

\subsubsection{Exoplanetary systems}
The correlation between long-range transport and the thermodynamical structure of the disk potentially signals a lack of small bodies on short-period orbits for the growing number of exoplanetary systems that host one or more inward-migratory planets (a result suggested in \citet{Batygin&Laughlin15}, but without a formal understanding of overstability). In other words, the inherently lower aspect ratio of circumstellar disks may prohibit long-range transport of smaller bodies by large planets that have migrated into the inner regions of their host disk. Moreover, as the hunt for exomoons continues to progress, understanding the limits to the migration of giant planet satellites will be key in unraveling their complexities.

\subsubsection{Within the Solar System}

Despite the existence of larger satellites orbiting exterior to numerous smaller satellites, the calculations carried out in this paper cannot be directly translated to the other satellite families of the Solar System. This is primarily because the evolution of the non-Jovian moons does not trace the simple primordial, inward-migration scenario that is generally envisioned for Jupiter's large satellites. Let us expand on this further.

In the cases of Uranus and Neptune, there are a number of distinctions, but first and foremost, formation of circumplanetary disks around sub-Jovian class planets prior to the disappearance of the protoplanetary nebula is unlikely \citep{Lambrechts+19,Krapp+24}. With respect to Uranus, the emergence of its satellite system is arguably best attributed to a giant impact that would have generated a debris disk from which the moons would have originated \citep{Morbidelli+12, Ida+20}. Regarding Neptune, the capture of Triton would have destroyed any pre-existing satellite system, and thus eliminated any of the primordial architecture \citep{Agnor&Hamilton06}. 

Now as for the Saturnian system, it does bear a closer resemblance to that of the Jovian, and it has even been demonstrated that satellite systems dominated by just a single large moon (Titan) can indeed form from giant planets' circumplanetary disks \citep{Fujii&Ogihara20}. Nonetheless, important differences persist. First, it is not clear that the magnetospheric cavity truncating the Saturnian disk would have been as robust as that of Jupiter's \citep{Batygin18,BatyginMorbidelli20}. So even if Saturn's Galilean analogue, Titan, originated at a period of a mere few days (as the works of \citealt{Fuller+16,Lainey+20,Goldberg&Batygin24} suggest; see however, \citealt{Jacobson22}), it is uncertain if the Saturnian equivalent to the Amalthea group would have even survived. Furthermore, the presence of Saturn's massive rings provides a second complicating factor, in that the small, under-dense inner moons likely possess a distinct dynamical evolution attributed to tidal disk spreading that gave birth to these massive rings (see e.g., \citealt{Crida&Charnoz12}). If, moreover, these rings are indeed young, it implies a relatively recent origin for the inner satellites that may have been freshly destabilized (e.g., \citealt{Wisdom+22,Cuk+24}), thus obscuring any primordial footprints of a long-range resonant transport scenario.

\subsection{Concluding remarks}

There is no doubt that Amalthea remains a puzzling piece in the larger picture of the Jovian system. Nonetheless, this tiny moon that remained unseen for centuries, always outshined by its more iconic neighbors, possesses valuable information that can be used to constrain the more general properties of its host environment. We thus hope that our scenario proposed here serves as a proper first step in quantifying the origin of Jupiter's small natural satellites, and in the end, a step towards a comprehensive theory for the late-stage evolution of our Solar System's largest planet.

\begin{acknowledgments}
\textit{Acknowledgments}.
IB is grateful to the Ahmanson Foundation for financial support, Hanno Rein and Dan Tamayo for help with \texttt{Rebound} implementation, along with Fred Adams, Matthew Belyakov, and Gabriele Pichierri for useful discussions. KB is thankful for the support of the David \& Lucile Packard Foundation, and the National Science Foundation (grant number: AST 2408867). Both authors thank Caltech and $^3$CPE. 
\end{acknowledgments}

\appendix
\section*{A brief note regarding capture}
\label{app:capture}

 Amidst the initial fanfare of Barnard's discovery, it became popular to assert that Amalthea is a captured object, likely coming from the asteroid belt, following a similar origin story trending for the then recently discovered Martian moons of thought comparable size and periodicity. \citet{Barnard92,Barnard93} himself, however, passionately opposed this capture hypothesis, again citing his careful measurements and the sensibly circularized, planar orbit of the small moon.

Despite his conviction born with some intuitive liberties, the capture hypothesis has lurked in the background of Amalthea's historical saga ever since, with the proposed reservoir for the nascent moon shifting along the way. The thought regained traction when the aforementioned \textit{Voyager} and \textit{Galileo} data motivated some to suggest that Amalthea ``accreted in the \textit{solar} nebula at or beyond Jupiter's position" (\citealt{Anderson+05}; emphasis ours), and was a product of ``capture from a heliocentric (asteroidal) orbit" \citep{Takato+04}, likely as a member of the Trojan asteroid family \citep{Prentice&terHaar79,Veverka+81,Prentice03}. We thus feel compelled to briefly address this idea and put it to rest. 

First, given Amalthea's circular coplanar orbit nestled between Io and Jupiter -- as noted by Barnard himself -- one should intuitively understand that a capture origin \textit{post}-nebular epoch (and by extension, \textit{post} the migration of the Galilean satellites) is not possible. As for during Jupiter's disk-bearing phase of existence, however, one could perhaps imagine the satellite being ``caught" by the Jovian nebula, then rapidly circularized from its hyperbolic orbit and delivered to its interior location all via the influence of gas drag alone -- and still yet, prior to the arrival of the Galilean satellites. To explain why this is not feasible, we must reflect on the effects of mass loss and ablation that arise when captured bodies encounter a medium of sufficient density. 

If an incoming planetesimal on the order $R \sim$ 100 km were somehow captured at a respectable semi-major axis of, say, $a\sim$ 40 $R_\text{J}$, with its natural hyperbolic orbit, it would result in near complete destruction of the small body -- with anything closer than that being even more violently destructive (recall Amalthea today at $a\sim 2.5\ R_{\rm{J}}$). In the study by \citet{RonnetJohansen20} -- who investigated the capture of planetesimals of $R =$ 1--100 km at $a >$ 40 $R_\text{J}$ in a notably thin, cool circumjovian disk -- they found that only $\sim$10\% of these captured objects will even survive ablation and remain at $a \gtrsim$ 10 $R_\text{J}$ (anything closer than 10 $R_\text{J}$ accretes onto Jupiter or is completely ablated). More importantly, however, of these surviving members, nearly 90\% are of $R <$ 10 km, with the largest being only $R \approx$ 65 km at approximately $a\sim30\ R_\text{J}$. 

Of course, one could continue to play with their imagination and propose a scenario in which a rather large planetesimal ($R >$ 100 km) were somehow captured \textit{well outward} of 40 $R_\text{J}$, and perhaps circularized amidst the damping of the disk before losing too much of itself to the surrounding medium. However, not only would this imply a starkly dense remnant (with volatiles being the first to go), even this rather contrived scenario would nonetheless still require a larger external perturber to subsequently deliver the stalled, circularized small body to a short-period orbit, seeing as the timescale for migration of this kind would be on the order of $\tau\sim 5 \times 10^4-10^5$ years (see \S \ref{sect:Just}).

Ergo, even in the unlikely event of capture by the circumjovian nebula, it would be dangerous and absurd to suggest that this capture process alone could result in the body coming to rest on a sensibly circularized, coplanar orbit at $a\sim 2.5\ R_\text{J}$ \textit{without} the eventual influence of the inward migrating Io as put forth in our scenario.

\bibliographystyle{aasjournal}
\bibliography{Amalthea}

\end{document}